\documentstyle[aps,prb,epsfig,preprint]{revtex}
\begin{document}
\draft
\title{On the "spin-freezing" mechanism in underdoped superconducting cuprates}
\author{M. Eremin$^{1,2}$, A. Rigamonti$^1$}
\address{Department of Physics " A. Volta", University of Pavia,\\
I-27100 Pavia, Italy}
\address{$^2$Physics Department, Kazan State University, 420008 Kazan, Russia}
\date{\today}
\maketitle

\begin{abstract}
The letter deals with the spin-freezing process observed by means of NMR-NQR
relaxation or by muon spin rotation in underdoped cuprate superconductors.
This phenomenon, sometimes referred as coexistence of antiferromagnetic and
superconducting order parameters, is generally thought to result from
randomly distributed magnetic moments related to charge inhomogeneities
(possibly stripes) which exhibit slowing down of their fluctuations on
cooling below T$_c$ . Instead, we describe the experimental findings as due
to fluctuating, vortex-antivortex, orbital currents state coexisting with
d-wave superconducting state. A direct explanation of the experimental
results, in underdoped Y$_{1-x}$Ca$_x$Ba$_2$Cu$_3$O$_{6.1}$ and La$_{2-x}$Sr$%
_x$CuO$_4$, is thus given in terms of freezing of orbital current
fluctuations.
\end{abstract}


%\twocolumn[\hsize\textwidth\columnwidth\hsize
%\csname @twocolumnfalse\endcsname

%]
%
\narrowtext
NMR-NQR and $\mu $SR experiments show that in underdoped cuprates, on
cooling from about Tc, divergent behaviour of the relaxation rates occur.
This phenomenon, generally interpreted in terms of ''glassy spin-freezing''
or of ''coexistence of superconductivity and magnetic ordering'', is
believed to be related to magnetic moments resulting from charge
inhomogeneties, possibly stripes. For $^{139}$La NMR-NQR observations in La$%
_{2-x}$Sr$_x$CuO$_4$, see Julien {\it et al.}\cite{1} and References
therein; a review of early data is given in Ref. \onlinecite{2}; the muon
longitudinal relaxation rate and spin precessional frequencies in La$_{2-x}$%
Sr$_x$CuO$_4$ and in Y$_{1-x}$Ca$_x$Ba$_2$Cu$_3$O$_6$ have been measured by
Niedermayer {\it et al.}\cite{3}. More recently the low-temperature
spin-freezing process in underdoped Y$_{1-x}$Ca$_x$Ba$_2$Cu$_3$O$_{6.1}$ ( T$%
_c\approx $ 35 K) has been studied by $^{89}$Y NMR relaxation in a field H$%
_0 $ = 9.4 T(Ref. \onlinecite{4}). The occurrence of a fast-relaxing
component in the $^{89}$Y spin-lattice relaxation below T$_c$ confirmed the $%
\mu $SR findings\cite{3} interpreted in terms of coexistence of
superconductivity and spin-freezing process. A new significant aspect was
the detection of the stretched character of the relaxation time\cite{4},
while the $\mu $SR data\cite{3} had been discussed in terms of a single
exponential recovery.

Very recently Mook {\it et al.}\cite{5} by means of inelastic neutron
scattering measurements in underdoped superconducting YBCO detected
longitudinal with respest to $c$-axis magnetic moments of unknown origin ,
with fluctuation frequencies well below the energy resolution of 1 meV.
Similar observation, but with transverse magnetic moments , was reported by
Sidis {\it et al.}\cite{6}. NMR-NQR and $\mu $SR spin-lattice relaxation data%
\cite{1,2,3,4} have been interpreted in terms of fluctuations of a local $%
{\bf h}(t)$ originating from magnetic moments associated to hole
localization \cite{7,8}or from staggered moments within locally ordered
mesoscopic domains \cite{9}, with non-zero effective spin\cite{10} or from
stripes\cite{11}. The microscopic origin of the fluctuating field ${\bf h}(t)
$ is still an open issue. Here we address this problem within a
substantially different picture , namely the one of the extended charge
density waves with imaginary order parameter ($id$-CDW). Important features
of this new scenario are described here and it is pointed out that the
fluctuating field originates from vortex- antivortex orbital currents.

We start from the so- called $t-J$ model by including the inter-site Coulomb
interaction: 
\begin{eqnarray}
H &=&\sum t_{ij}\psi _i^{pd,\sigma }\psi _j^{\sigma ,dp}+\frac 12\sum
J_{ij}\left[ \left( S_iS_j\right) -\frac{n_in_j}4\right] +  \nonumber \\
&&{}+\frac 12\sum G_{ij}\delta _i\delta _j  \label{eq:1}
\end{eqnarray}
with $\delta _i$ being the number of extra holes per unit cell of bilayer, $%
\psi _i^{pd,\sigma }$, ($\psi _j^{\sigma ,dp}$) creation (annihilation)
operator constructed on the basis of the singlet combination of copper ($d$)
and oxygen states ($p$) ( see Ref. \cite{12} for details). It should be
noted that the temperature dependence of Cu(2) Knight shift in YBa$_2$Cu$_4$O%
$_8$, the evolution of the Fermi surface in Bi$_2$Sr$_2$CaCu$_2$O$_{8+y}$
and the {\bf k}-dependence of the pseudogap in the normal state have been
successfully explained \cite{13} in the framework of that model, including
the $d$-wave orbital symmetry of the superconducting state (SC) \cite{14}.
The $d$-wave SC can coexist with extended charge density waves (for short $%
id $-CDW) \cite{15,16}. Chakravarty {\it et al.} \cite{17} have stressed
that the corresponding ordered phase is a staggered pattern of orbital
currents. The region in which $d$-wave SC coexists with $id$-CDW strongly
depends on doping \cite{15}. For $\delta _{opt}$ the SC phase forces out $id$%
-CDW state (see Fig. 1 in Ref. \onlinecite{15}), but when T$^{*}$/T$_c\leq $%
2.5 then SC and $id$-CDW state can coexist even at low temperatures.
Non-monotonic temperature dependence of energy gap parameter, due to the
competition of SC and $id$-CDW, was predicted in Ref. \cite{15} and
confirmed by Ekino {\it et al.} \cite{18} by means of break-junction
tunneling spectroscopy in Bi$_2$Sr$_2$CaCu$_2$O$_{8+y}$. Starting from
another form of Hamiltonian Murakami \cite{19} has predicted (for $G<0$) the
coexistence $id$-CDW, antiferromagnetic (AF) and $\eta $-triplet states near
half filling. Coexistence of d-CDW and spin density waves (SDW) was also
derived by Bouis {\it et al.} \cite{20}.

The main issue related to the scenario recalled above is whether it is
possible to have a coexistence of $d$-SC and AF or SDW in the frame of the
singlet - correlated band Hamiltonian (\ref{eq:1}). Using the Roth - type
decoupling scheme \cite{21}, in a mean field approximation, from Eq.(1) one
has 
\begin{equation}
H^{MF}=\sum_k\Psi _k^{+}M_k\Psi _k  \label{eq:2}
\end{equation}
where the matrix $M_k$ is 
\begin{equation}
M_k=\left( 
\begin{array}{cccc}
\varepsilon _k-\mu & \eta _{k,Q}^{\uparrow } & \Delta _k^{\uparrow } & 
U_{k,Q}^{\uparrow } \\ 
\eta _{k+Q,-Q}^{\uparrow } & \varepsilon _{k+Q}-\mu & U_{k+Q,-Q}^{\uparrow }
& \Delta _{k+Q}^{\uparrow } \\ 
-(\Delta _{-k}^{\downarrow })^{*} & -(U_{-k,-Q}^{\downarrow })^{*} & 
-\varepsilon _{-k}+\mu & -(\eta _{-k,-Q}^{\downarrow })^{*} \\ 
-(U_{-k-Q,Q}^{\downarrow })^{*} & -(\Delta _{-k-Q}^{\downarrow })^{*} & 
-(\eta _{-k-Q,Q}^{\downarrow })^{*} & -\varepsilon _{-k-Q}+\mu
\end{array}
\right)  \label{eq:3}
\end{equation}
Let us focus now at the order parameters: 
\begin{eqnarray}
\eta _{k,Q}^{\uparrow } &=&G^{ph}+\left[ t_{k+Q}-\frac{\left( \left\langle
s_is_j\right\rangle t_{ij}\right) _k}{(P_{pd})^2}\right] \left\langle \frac 1%
2e_Q+s_Q^z\right\rangle -  \nonumber \\
&&{}-\frac 1{2NP_{pd}}\sum_{k^{\prime }}\left\{ J_{k^{\prime
}-k}\left\langle \psi _{k^{\prime }+Q}^{pd,\downarrow }\psi _{k^{\prime
}}^{\downarrow ,dp}\right\rangle +2G_{k^{\prime }-k}\left\langle \psi
_{k^{\prime }+Q}^{pd,\uparrow }\psi _{k^{\prime }}^{\uparrow
,dp}\right\rangle \right\}  \label{eq:4}
\end{eqnarray}
and 
\begin{eqnarray}
U_k^{\uparrow } &=&-\frac 1{NP_{pd}}\sum_{k^{\prime }}\left( t_{k^{\prime
}+Q}+t_{k^{\prime }}\right) \left\langle \psi _{k^{\prime }}^{\uparrow
,dp}\psi _{-k^{\prime }-Q}^{\downarrow ,dp}\right\rangle +  \nonumber \\
&&{}+\frac 1{2NP_{pd}}\sum_{k^{\prime }}\left[ J_{k^{\prime
}-k}+J_{k^{\prime }+k+Q}-2G_{k^{\prime }-k}\right] \left\langle \psi
_{k^{\prime }}^{\uparrow ,dp}\psi _{-k^{\prime }-Q}^{\downarrow
,dp}\right\rangle  \label{eq:5}
\end{eqnarray}
where $t_k$ ,$J_k$, $G_k$ are the Fourier transforms of transfer,
superexchange and Coulomb coupling parameters, respectively. $e_Q$ and $%
s_Q^z $ are the Fourier amplitudes of the conventional charge density waves
and spin density waves. For the commensurate instability wave vector {\bf Q}
=($\pi ,\pi $) as it was observed in neutron scattering \cite{5,6} one
derives the following relations 
\begin{eqnarray}
\eta _{k,Q}^{\uparrow } &=&S_{1k}^{\uparrow }+S_{2k}^{\uparrow
}+iD_k^{\uparrow }\qquad  \nonumber \\
\eta _{k+Q,-Q}^{\uparrow } &=&-(S_{1k}^{\uparrow })^{*}+(S_{2k}^{\uparrow
}+iD_k^{\uparrow })^{*}\qquad  \nonumber \\
(\eta _{-k,-Q}^{\downarrow })^{*} &=&S_{1k}^{\downarrow }+S_{2k}^{\downarrow
}-iD_k^{\downarrow }\qquad  \nonumber \\
(\eta _{-k-Q,Q}^{\downarrow })^{*} &=&-(S_{1k}^{\downarrow
})^{*}-(S_{2k}^{\downarrow }-iD_k^{\downarrow })^{*}  \label{eq:6}
\end{eqnarray}
Let us consider now the quantities 
\begin{equation}
S_{1k}^{\uparrow }=\left[ t_{k+Q}^{(1)}-\frac{\left( \left\langle
s_is_j\right\rangle t_{ij}^{(1)}\right) _k}{(P_{pd})^2}\right] \left\langle 
\frac 12e_Q+s_Q^z\right\rangle  \label{eq:7}
\end{equation}
and 
\begin{equation}
S_{2k}^{\uparrow }=G^{ph}+\left[ t_{k+Q}^{(2)}-\frac{\left( \left\langle
s_is_j\right\rangle t_{ij}^{(2)}\right) _k}{(P_{pd})^2}\right] \left\langle 
\frac 12e_Q+s_Q^z\right\rangle  \label{eq:8}
\end{equation}
where the indexes 1 and 2 refer to the first and the second nearest
neighbour hopping parameters. $S_{1k}^{\uparrow }$ changes the sign under
the transformation ${\bf k}\longrightarrow {\bf k+Q}$ and as one can see
from Eq. (\ref{eq:3}) this yields a damping factor, since the energy
dispersion becomes complex. The value $S_{2k}^{\uparrow }$ is invariant
under such transformation and can generate $s$-wave CDW and SDW. Therefore,
for open Fermi surface one finds that at the hopping parameter t$_1$ larger
enough the expectation values $<e_q>$ and $<s_q>$ must be zero, and
conventional SDW/CDW do not exist. This conclusion is in agreement with the
one derived in Ref. \cite{22} on the basis of neutron scattering data.
However, we stress that the above conclusion is valid only for the energy
dispersion of the form ($\cos k_x+\cos k_y$), but not for the case $\epsilon
_k=(\cos k_x+\cos k_y)^2$ when pockets are formed around the points ($\pm
\pi /2,\pm \pi /2$) in the Brillouin zone. In our picture this happens when
antiferromagnetic correlation (parameter $J$) hampers the nearest neighbour
hopping {\it i. e.} $t_1\left( P_{pd}+\frac{\left\langle s_is_j\right\rangle 
}{P_{pd}}\right) \approx 0$. It is noted that the order parameter $U_k$
resembles the so-called $\eta $-singlet pairing introduced by Yang \cite{23}%
. Formally the $U_k$-order parameter construction in Eq.(\ref{eq:5}) also
resembles the Larkin-Ovchinnikov, Fulde-Ferrell parameter \cite{24,25} with
the important difference that the instability wave vector {\bf q} is
completely different. The leading term of $U_k$ 
\begin{equation}
U_0^{\uparrow }=-\frac 1{NP_{pd}}\sum_{k^{\prime }}\left( t_{k^{\prime
}+Q}+t_{k^{\prime }}\right) \left\langle \psi _{k^{\prime }}^{\uparrow
,dp}\psi _{-k^{\prime }-Q}^{\downarrow ,dp}\right\rangle  \label{eq:9}
\end{equation}
is determined by the second- and third- neighbours hopping integrals.
Numerical solution of Eq. (\ref{eq:9}) in correspondence to the parameters ($%
t_1$=80 meV, $t_2$ = 0 and $t_3$= 12 meV) used \cite{12} for the Fermi
surface near optimal doping, yields a critical temperature of the onset $U$%
-phase below 1 K. One should remark that the onset of this inhomogeneous
superconducting state is very sensitive to magnetic impurities and to the
details of the Abrikosov's vortex lattice. In addition we note that the $U$%
-ordering is particle-particle pairing at opposite spin orientations and
must be sensitive to the external magnetic field. Since there is no evidence
of such an effect of the magnetic field (see the data later on) we conclude
that the peak in relaxation rate is not related to phase transition
involving this $U$ order parameter. In the matrix given by Eq. (\ref{eq:3})
there are two imaginary components $iD^{\uparrow }$ and $iD^{\downarrow }$.
It is useful to introduce their combinations 
\begin{equation}
i(D_{k,Q}^{\uparrow }+D_{k,Q}^{\downarrow })=-\frac 1{2NP_{pd}}%
\sum_{k^{\prime }}\left( J_{k^{\prime }-k}+2G_{k^{\prime }-k}\right) \left\{
\left\langle \psi _{k^{\prime }+Q}^{pd,\downarrow }\psi _{k^{\prime
}}^{\downarrow ,dp}\right\rangle +\left\langle \psi _{k^{\prime
}+Q}^{pd,\uparrow }\psi _{k^{\prime }}^{\uparrow ,dp}\right\rangle \right\}
\label{eq:10}
\end{equation}
\begin{equation}
i(D_{k,Q}^{\uparrow }-D_{k,Q}^{\downarrow })=-\frac 1{2NP_{pd}}%
\sum_{k^{\prime }}\left( J_{k^{\prime }-k}-2G_{k^{\prime }-k}\right) \left\{
\left\langle \psi _{k^{\prime }+Q}^{pd,\downarrow }\psi _{k^{\prime
}}^{\downarrow ,dp}\right\rangle -\left\langle \psi _{k^{\prime
}+Q}^{pd,\uparrow }\psi _{k^{\prime }}^{\uparrow ,dp}\right\rangle \right\}
\label{eq:11}
\end{equation}
the first corresponding to charge current and the second one to spin current 
\cite{26,27} or, in other terminology, to the spin-nematic state \cite{28,29}%
. It is clear from Eqs. (\ref{eq:10})-(\ref{eq:11}) that $T^{cr}(id-$CDW$%
)>>T^{cr}(id-$SDW$)$, and when $2G\approx J$ Eq. (\ref{eq:11}) has no
solution at all. Concluding this theoretical analysis, we see that the real
situation is strongly dependent on the Fermi surface. Because there are no
indication about pockets-like Fermi-surface in superconducting underdoped
cuprates, we do expect coexistence of the $d$-SC, $id$-CDW and $id$-SDW or $%
U $-state at low temperatures. Let us see now what the experimental findings
tell us .

On general physical grounds NMR-NQR or muon relaxation rates can be written 
\begin{equation}
1/T_1\approx \gamma _n^2\left\langle h_{eff}^2\right\rangle J\left( \omega
_m,\tau _e\right)  \label{eq:12}
\end{equation}
where $\left\langle h_{eff}^2\right\rangle $ is a mean square amplitude of
the transverse effective field at the nuclear or muon site and $\left(
\omega _m,\tau _e\right) $ is the spectral density of the fluctuations at
the measuring frequency $\omega _m$ and an effective correlation time $\tau
_e$ is assumed. The stretched exponential recovery, of the form $\exp \left[
-t/T_1^e\right] ^{1/2}$ observed experimentally \cite{1,4} is naturally
explained by charge density waves scenario, as it was pointed out by Philips 
\cite{30}. Usually this process is described as a superposition of Debye
relaxation or in an ideal limit as a Laplace transform 
\begin{equation}
\exp \left[ -\frac t{T_1^e}\right] ^{1/2}=\int_0^\infty \rho (1/T_1)\exp
\left[ -t/T_1\right] d(1/T_1)  \label{eq:13}
\end{equation}
where $\rho (1/T_1)=\frac{T_1}{2\sqrt{\pi }}\left( \frac{T_1}{T_1^e}\right)
^{1/2}\exp \left[ -\frac{T_1}{4T_1^e}\right] $ is the distribution function.
It is conceivable to assume that the distribution of (1/T$_1$) is related to
a distribution of correlation time $\tau _e$ and in turn to a distribution
of the energy barrier $E$ pinning the fluctuating - sliding current motions.
Then we write 
\begin{equation}
\left\langle \tau _e\right\rangle =\tau _0\exp (\left\langle E\right\rangle
/k_BT)  \label{eq:14}
\end{equation}
and 
\begin{equation}
J\left( \omega _m,\tau _e\right) =\frac{2\left\langle \tau _e\right\rangle }{%
1+\omega _m^2\left\langle \tau _e\right\rangle ^2}  \label{eq:15}
\end{equation}
When the progressive slowing-down , on cooling, of the orbital current
excitations causes $<\tau _e>\approx \omega _m^{-1}$, one has the maximum in
the relaxation rates (frequently used to define $T_g$, commonly called ''
spin -glass'' freezing temperature). From this maximum one can extract the
effective magnetic field induced by the currents at the nuclear or muon
site: 
\begin{equation}
\sqrt{\left\langle h_{eff}^2\right\rangle }=\frac 1{\gamma _n}\left[ \frac{%
\omega _m}{\left( T_1^e\right) _{\max }}\right] ^{1/2}  \label{eq:16}
\end{equation}
One could remark that on the basis of the above equation a slight
underestimate of $\sqrt{\left\langle h_{eff}^2\right\rangle }$ is obtained,
since the distribution of correlation times tends to level the maximum of
the spectral function. The distribution also accounts from the deviation of
the experimental data from the simple law ( Eq. (\ref{eq:14})) in the low
temperature region (see Fig. 1). The values of $\sqrt{\left\langle
h_{eff}^2\right\rangle }$ derived from the experimental data on the basis of
Eq. (\ref{eq:16}) are collected in Table 1.

In Fig. one sees that Eq. (\ref{eq:12}), with Eqs. (\ref{eq:14})-(\ref{eq:15}%
) rather well fits the observed temperature dependence of relaxation time in
spite of the roughness of the assumptions. The energy barrier $<E>$ is
roughly the same, for different kind of experiments and systems (see also
Table 1). It is naturally related to periodic lattice potential. It should
be stressed that the energy barrier is almost {\it independent on the
applied magnetic field}. This fact would be hard to explain in the
spin-freezing scenario. On the contrary, in our picture the correlation
functions entering Eq.(\ref{eq:10}) are of particle - hole type and they are
proportional to difference of Fermi functions with the same spin orientation 
\cite{13}. Therefore, Zeeman energy comes out from these functions and they
become independent of the external magnetic field. As a result the pseudo-
gap temperature $T^{*}$ and the pinning process caused by the lattice
periodic potential are insensitive to the magnetic field.

As one can see from Table that the order of magnitude $\sqrt{\left\langle
h_{eff}^2\right\rangle }$ of the magnetic field agrees with theoretical
estimation made on the basis of similar bond currents patterns (flux-phase)
Ref. \cite{32}.

The $z-$axis of the electric field gradient at $^{139}La$ site is almost
parallel to $c-$axis ( see for Ref. \onlinecite{2}). Therefore we can
interpret $\sqrt{\left\langle h_{eff}^2\right\rangle }=$255Gs as a field
transeverse with respect to $c$-axis. The effective field at $Y$ site (9Gs)
is relatively small. This means that transeverse components within $Cu-O$
bilayer are almost compensated. Two magnetic field components at nuclear
spin occur: a direct component - $h(1)$ as dicussed above and so-called
hyperfine and supertransferred fields from adjacent Cu(2) states - $h(2)$.
The $id$-CDW state contributes to both. The modulation of the hyperfine
coupling yields the effective magnetic field which is proportional to
charge-spin correlation function 
\begin{equation}
h(2)\propto F(q-q^{\prime })e_qS_{q^{\prime }}  \label{eq:17}
\end{equation}
where $F(q)$ is the form factor involved in the Mila-Rice Hamiltonian (see
Ref. \cite{2}). This factor is different for $La$ and $Y$ nuclei sites and
for electronic shell of ions $Yb^{3+}$ and $Er^{3+}$ as well.

In summary , we have described the so- called spin-freezing process in
superconducting state of underdoped cuprates with non-zero first neighbours
hopping in a frame of a regime of coexistence of $d-$SC and $id-$CDW states .

We have shown:

i) that this scenario is naturally derived in a frame of singlet -
correlated band model (or on the basis of the t-J model including inter
-site Coulomb repulsion) and it is consistent with recent neutron scattering
data and tunneling spectroscopy measurements. ii) the progressive slowing
down of the sliding current motions, due to pinning barrier related to
periodic lattice potential explains the temperature behavior of the
relaxation rates detected in NMR-NQR, $\mu $SR and EPR experiments; in
particular it directly justifies the stretched exponential character of the
recovery curves; the fluctuating frequency lying in 10-1000 MHz range
explain why the resolution limit ($\approx $ meV) makes them invisible in
neutron scattering measurements iii) the fact that practically no difference
is observed in the relaxation rates upon increasing the magnetic field from
zero $(NQR)$ up to 23 Tesla and when the field is applied parallel or
perpendicular to the c- axis ( that is not explained in a spin-freezing
scenario ) is explained by the insensitivity of the current correlation
functions to the external magnetic field .

Thus we believe that our picture brings a new and suggestive perspective,
which is fully supported by the experimental findings and reveals new
insights on a much debated issue.

Useful discussions with R. Zeyher, P. Carretta, M. Corti , A. Dooglav ,I.
Eremin and F. Tedoldi are acknowledged. M. V. Eremin is grateful to
Landau-Network-Centro - Volta foundation for the fellowship and possibility
to stay at University of Pavia.His work in Russia was supported by Swiss
National Science Foundation (Grant N 7SUPJ062258) and Russan Program ''
Superconductivity'' (Project N 98014-1). The work is a part of the project
of advanced recearch ( PRA - SPIS) of INFM.

\begin{figure}[h]
%
%\vspace{-1.1cm}
%
%\centerline{\epsfig{clip=,file=Cortibw1.eps,width=7.5cm,angle=-90}}
%\vspace{-0.75cm}
\centerline{\epsfig{clip=,file=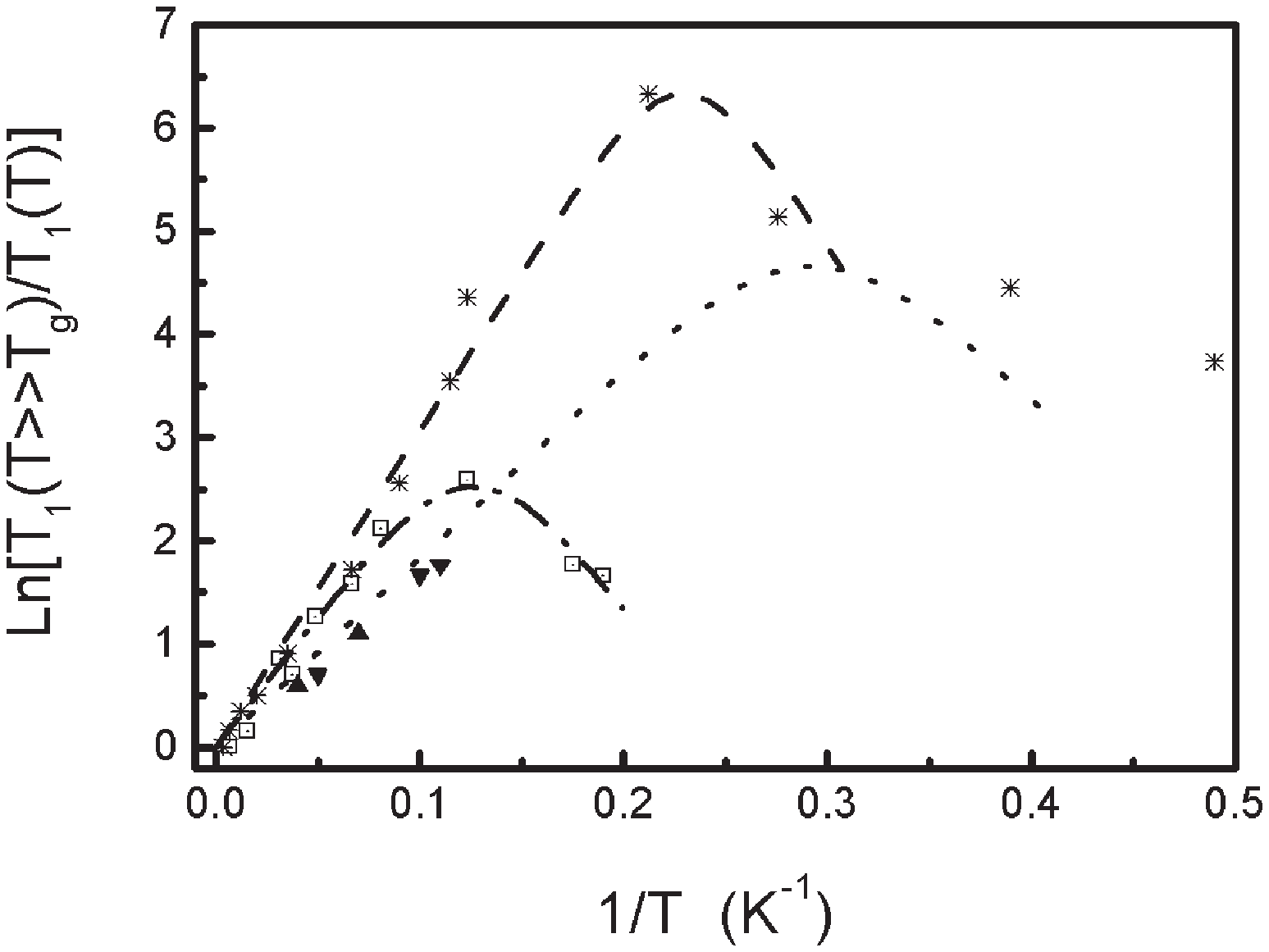,width=15.5cm,angle=0}} \vspace{0.5ex}
\caption{ Temperature behavior of the relaxation rates in La$_{2-x}$Sr$_x$CuO%
$_4$ and in Y$_{1.85}$Ca$_{0.15}$Ba$_2$Cu$_3$O$_{6.1}$, normalized to the
values at T$>>$T$_g \approx$ T$_{max}$, as a function of T$^{-1}$. The data
are taken from Ref. \cite{1,4}: circles are for La$_{1.94}$Sr$_{0.06}$CuO$_4$
in zero field and resonance frequency 18.6 MHz (NQR), stars - La$_{1.9}$Sr$%
_{0.1}$CuO$_4$ in H=23.2 Tesla, parallel to the c-axis, uptriangles - La$%
_{1.9}$Sr$_{0.1}$CuO$_4$ with H=9.4 Tesla, perpendicular to the c-axis,
downtriangles - La$_{1.9}$Sr$_{0.1}$CuO$_4$ at H=9.4 Tesla, parallel to the
c-axis, squares - Y$_{1.85}$Ca$_{0.15}$Ba$_2$Cu$_3$O$_{6.1}$ in H=9.4 Tesla. 
}
\label{fig3}
\end{figure}
\begin{table}[tbp]
{{\bf Table 1} Maximum values of the relaxation rates, estimated values for
the effective field and for the average energy barrier for pinning of the
sliding motions of the orbital currents.}\\[1ex]
\begin{tabular}{|c|c|c|c|c|c|c|}
{\bf sample} & {\bf Method (Ref.)} & {\bf $\nu_0$(MHz)} & {\bf H$_0$(Tesla)}
& {\bf 1/T$_1$(s$^{-1}$)} & {\bf $<$ E $>$(K)} & {\bf $\sqrt{<h_0^2>}$(Gs)}
\\ \hline
La$_{1.9}$Sr$_{0.1}$CuO$_4$ & $^{139}$La NMR \cite{1} & 139.528 & 23.2 & $%
\approx$700 & 19 & 210 \\ 
La$_{1.94}$Sr$_{0.06}$CuO$_4$ & $^{139}$La NQR \cite{1} & 18.6 & 0 & $%
\approx $8000 & 30 & 255 \\ 
Y$_{0.85}$Ca$_{0.15}$Ba$_2$Cu$_3$O$_{6.1}$ & $^{89}$Y NMR \cite{4} & 19.61 & 
9.4 & $\approx$1 & 26 & 9 \\ 
Y$_{1-x}$Ca$_{x}$Ba$_2$Cu$_3$O$_{6.02}$ & $\mu$SR \cite{3} & 0.51 & 0 & $%
\approx$60000 & 0 & 150 \\ 
YBa$_2$Cu$_3$O$_{6.85}$ & Er,Yb EPR \cite{30} & 9460 & 0.2 & - & 25 & 160 \\ 
&  &  &  &  &  & 
\end{tabular}
\end{table}

\end{document}